\documentclass[aps,prc,twocolumn,showpacs,superscriptaddress,footinbib]{revtex4-1}

\usepackage{graphicx}
\usepackage{dcolumn}
\usepackage{bm}
\usepackage{enumerate}

\usepackage[normalem]{ulem}  
\usepackage[dvips]{color}

\begin{document}

\title{Influence of secondary decay on odd-even staggering of fragment cross sections}

\author{J.R. Winkelbauer}
\affiliation{National Superconducting Cyclotron Laboratory and Department of Physics and Astronomy Department,\\ Michigan State University, East Lansing, Michigan 48824, USA}
\author{S.R. Souza}
\affiliation{Instituto de F\'\i sica, Universidade Federal do Rio de Janeiro
Cidade Universit\'aria, \\CP 68528, 21941-972, Rio de Janeiro, Brazil}
\affiliation{Instituto de F\'\i sica, Universidade Federal do Rio Grande do Sul,\\
Av. Bento Gon\c calves 9500, CP 15051, 91501-970, Porto Alegre, Brazil}
\author{M.B. Tsang}
\affiliation{National Superconducting Cyclotron Laboratory and Department of Physics and Astronomy Department,\\ Michigan State University, East Lansing, Michigan 48824, USA}

\date{\today}

\begin{abstract}

Odd-Even Staggering (OES) appears in many areas of nuclear physics, and is generally associated with the pairing term in the nuclear binding energy. To explore this effect, we use the Improved Statistical Multifragmentation Model to populate an ensemble of hot primary fragments, which are then de-excited using the Weisskopf-Ewing statistical emission formalism. The yields are then compared to experimental data. Our results show that, before secondary decay, OES appears only in the yields of even mass fragments and not in the yields of odd mass fragments. De-excitation of the hot fragments must be taken into account to describe the data, suggesting that the OES in fragment yields is a useful criterion for validating or adjusting theoretical de-excitation models.

\end{abstract}

\maketitle

\begin{section}{Introduction}
\label{sect:introduction}

Odd-Even Staggering (OES) is a widely observed phenomenon in nuclear physics. 
The early observed OES in nuclear masses associate the phenomenon with the pairing term in the binding energy \cite{neutronProtonPairingFriedmanBertsch2007,OddEvenFriedmanBertsch2009}. 
Since particle production yields, including fission fragments, correlate strongly with the binding energy, measured fragment cross sections exhibit OES effects.  
Sawtooth-shaped charge correlations, based on the $Z$ distributions of fragments produced in different reaction mechanisms including fission and multifragmentation, have been reported \cite{staggeringRicciardi,staggeringDAgostino,staggeringCasini}. 
Much of the previous work focused on staggering as a function of the atomic number, $Z$, because most experimental data was limited to only elemental identification. 

Recently, isotopic identification for heavy reaction products can be achieved with state-of-the-art detectors. 
When the charge distributions are subdivided according to the neutron excess of the fragments, the staggering plots with isotope resolution reveal a more complex structure suggesting the de-excitation of the hot fragments contribute significantly to the observed OES effect. 
In fragments that are very neutron-rich, the OES effect may be reversed, i.e. production of odd mass fragments are enhanced, compared to the less n-rich fragments. 
In this work, we investigate the OES effects of the isotopic fragment distributions obtained in the projectile fragmentation of $^{40}$Ca, $^{48}$Ca, $^{58}$Ni, and $^{64}$Ni at 140 MeV/nucleon whose experimental analysis has been reported in Ref.\ \cite{projFragMSU2006}.

Using the grand-canonical version of the Improved Statistical Multifragmentation Model (ISMM) \cite{ISMMmass} and the Weisskopf-Ewing statistical emission to describe the de-excitation of the hot primary fragments, we examine the effects associated with the latter on the observed staggering.
Our results suggest that the OES effect can be very useful in constraining the treatments adopted in the description of the de-excitation process. 
The remainder of the manuscript is organized as follows.
In Sec.\ \ref{sect:theory} we briefly recall the main features of the ISMM.
The results are presented and discussed in Sec.\ \ref{sect:results}, and the conclusions are presented in Sec.\ \ref{sect:conclusions}.

\end{section}

\begin{section}{Theoretical Framework}
\label{sect:theory}
In the framework of the grand-canonical approach \cite{isoMassFormula2008,finiteSizeIsoRatios}, the yields $Y(A,Z)$ of a fragment with mass and atomic numbers, respectively, $A$ and $Z$, reads:

\begin{equation}
\label{eq:ay}
Y(A,Z)=\frac{g_{A,Z}V_fA^{3/2}}{\lambda_T^3}e^{-[f^{\rm smooth}_{A,Z}(T)-B^{\rm pairing}_{A,Z}-\mu_pZ-\mu_b A]/T}\\
\end{equation}

\noindent
where $g$ stands for the spin degeneracy (taken as unit, except for the empirical values used for $A\le 4$),  $\lambda_T=\sqrt{2\pi\hbar^2/m_nT}$ is the thermal wavelength, $m_n$ denotes the nucleon mass, $\mu_p$ ($\mu_b$) represents the proton (baryon) chemical potential, $f^{\rm smooth}_{A,Z}(T)$ is the Helmholtz free energy associated with the fragment plus the pairing term of the binding energy $B^{\rm pairing}_{A,Z}$.
The free volume reads $V_f=\chi V_0$, where $V_0$ is the source's volume at normal density and we use $\chi=2$ throughout this work.
As in Refs.\ \cite{isoMassFormula2008,finiteSizeIsoRatios}, the Helmholtz free energy has contributions from the fragment's binding energy $B_{A,Z}$, terms associated with the Wigner-Seitz correction \cite{smm1} to the Coulomb energy $C_c\frac{Z^2}{A^{1/3}}\frac{1}{(1+\chi)^{1/3}}$, besides those related to the internal excitation of the fragment $f^*_{A,Z}$:

\begin{eqnarray}
\label{eq:fsmooth}
f^{\rm smooth}_{A,Z}(T)&=&f^*_{A,Z}(T)-[B_{A,Z}-B^{\rm pairing}_{A,Z}]\\
&-&C_c\frac{Z^2}{A^{1/3}}\frac{1}{(1+\chi)^{1/3}}\;.\nonumber
\end{eqnarray}

\noindent
For clarity, we use the standard SMM parameterization of the internal free energy \cite{isoMassFormula2008}, which is isospin independent, {\it i.e.}, $f^*_{A,Z}(T)=f^*_A(T)$.
The pairing term is written as:

\begin{equation}
\label{eq:bpairing}
B^{\rm pairing}_{A,Z}=(-1)^Z\frac{\delta_p}{A^{\tau}}\frac{1}{2}[1+(-1)^{N-Z}]\;,
\end{equation}

\noindent
where $N=A-Z$ denotes the neutron number, and $\delta_p$ is the pairing energy.
In this work we employ the liquid drop mass formula presented in Ref.\ \cite{ISMMmass}, and we refer the reader to that work for the numerical values of the parameters. Although we have employed a liquid drop parametrization of the binding energy in all the calculations, we have checked that our conclusions are not affected by the use of more precise values of the binding energy, such as those given by the procedure described in Ref.\ \cite{ISMMlong}.

With these definitions, $Y(A,Z)$ may be rewritten as the product of a smooth term multiplied by a rapidly oscillating function of $Z$:

\begin{equation}
\label{eq:ysmoothFluct}
Y(A,Z)=Y_{\rm smooth}(A,Z)\times e^{(-1)^Z\frac{\delta_p}{A^\tau}\frac{1}{2T}[1+(-1)^{N-Z}]}\;,
\end{equation}

\noindent
where

\begin{equation}
\label{eq:ysmooth}
Y_{\rm smooth}(A,Z)=e^{-[f^{\rm smooth}_{A,Z}(T)-\mu_p Z-\mu_b A]/T}\;.
\end{equation}

The chemical potentials are determined by simultaneously solving the equations:

\begin{equation}
\label{eq:sumA}
A_0=\sum_{A,Z}A\;Y(A,Z)
\end{equation}

\noindent
and

\begin{equation}
\label{eq:sumZ}
Z_0=\sum_{A,Z}Z\;Y(A,Z)
\end{equation}

\noindent
where $A_0$ ($Z_0$) denotes the mass (atomic) number of the decaying source.

The de-excitation of the hot primary fragments is taken into account through the Weisskopf-Ewing statistical emission, as described in Refs.\ \cite{ismmFlow2007,isoSMMTF}.
We consider the emission of nuclei up to $^{20}$O.
For consistency, the same values of the binding energy $B_{A,Z}$ are also used in the decay treatment.
The same remarks hold for the density of excited states, as explained in Ref.\ \cite{ismmFlow2007}.

\end{section}

\begin{section}{Results}
\label{sect:results}
The experimental Z distributions obtained in the projectile fragmentation of $^{40}$Ca, $^{48}$Ca, $^{58}$Ni, and $^{64}$Ni at 140 MeV/nucleon on $^9$Be reported in Ref.\ \cite{projFragMSU2006} are shown in Fig.\ \ref{fig:zDist}, where the normalized quantity 

\begin{equation}
\label{eq:yzNorm}
Y(Z)=\sum_A Y(A,Z)/\sum_{A,Z} Y(A,Z)\;,
\end{equation}

\noindent
is plotted for each case.
The data, represented by the full circles, increases, on the average, as a function of the fragment's atomic number $Z$. A detailed inspection reveals that $Y(Z)$ deviates up and down from the local average value.
As observed before, this staggering is weaker in the reactions with more neutron-rich projectiles.

\begin{figure}[tbh]
\includegraphics[width=8.5cm,angle=0]{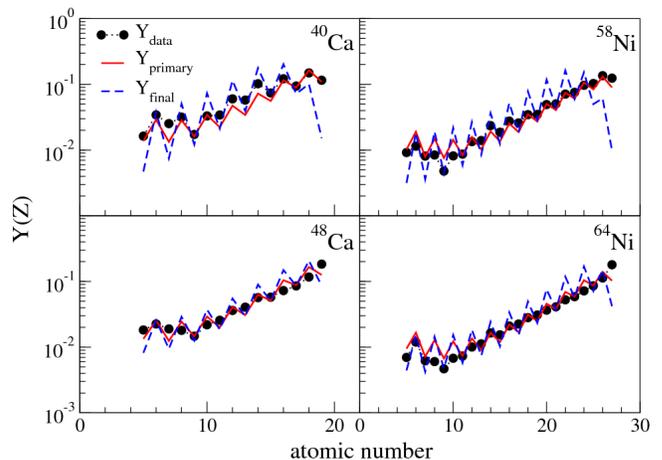}
\caption{\label{fig:zDist} (Color online) Z distribution of fragments observed in the projectile fragmentation of several projectiles on a $^9$Be target.
The data are taken from Ref.\ \cite{projFragMSU2006} whereas the calculations correspond to the grand-canonical version of the ISMM, using the liquid drop formula of Ref.\ \cite{ISMMmass}, at breakup temperature $T=4.0$~MeV.  For details, see the text.}
\end{figure}

The predictions made by the ISMM, presented in Sec.\ \ref{sect:theory}, are also displayed in this figure and are represented by the full (primary yields) and dashed (final yields) lines. There is no attempt to fit the data with the calculation. The model is primarily used to give insights into the OES. Instead of doing calculations with different values of temperature in order to account for the variation of the centrality of the collisions, for clarity and simplicity, we adopted a single average value of $T$ for all systems.
The primary distribution was generated using $T=4.0$~MeV.
This choice was based on the criterion that the yields, after the de-excitation of the primary fragments, would follow the experimental distribution as close as possible.
Slightly different average temperatures could have been used, but this would not affect the conclusions of this paper.
The model results follow the experimental trend of exhibiting weaker staggering effects in the case of the neutron rich projectiles.
Furthermore, by comparing the primary and final distributions, one sees that the deviations are clearly enhanced by the decay of the excited fragments.

To isolate the local staggering behavior, following Ref.\ \cite{staggeringCasini}, we obtain the average value $\overline{Y}_{Z}(Z)$ by carrying out a parabolic fit, considering two points to the left and to the right of $Z$, besides its own value, except when $Z$ is close to or at either ends of the $Z$ range, where the points lying beyond the data edge are obviously not used. More precisely, although five points are always employed in the fit, one shifts the selected region so that it does not extend beyond the edges. The ratio:
\begin{equation}
\label{eq:ratio}
R_{Z}(Z)=Y_{Z}(Z) / \overline{Y}_{Z}(Z)
\end{equation}
is plotted in Fig. \ref{fig:OESz}, for the data from Ref. \cite{projFragMSU2006} shown in Fig. \ref{fig:zDist}. The features of this figure are consistent with previous studies. There is a weaker effect for the neutron-rich projectile, and the trend generally decreases with increasing Z. 
\\
\\
\begin{figure}[tbh]
\includegraphics[width=8.5cm,angle=0]{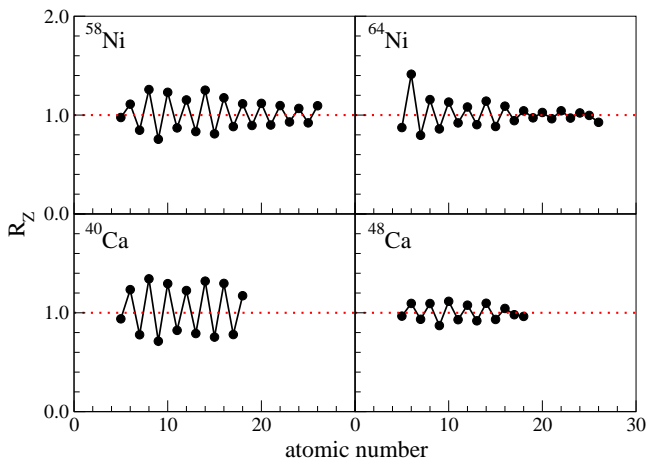}
\caption{\label{fig:OESz} (Color online)  Ratio between $Y_{Z}(Z)$ and $\overline{Y}_{Z}(Z)$ for the four projectiles in Ref. \cite{projFragMSU2006}.}
\end{figure}

In the literature, the staggering effects have been associated with the pairing energy, $\delta_p$. According to Eq. \ref{eq:ysmoothFluct}, the staggering effects would be different depending on whether the neutron excess, $N-Z$, is even or odd. In the ISMM framework, the only non-smooth contribution to the primary yields is the empirical binding energies. This is illustrated in Fig. \ref{fig:58NiBE} for the $^{58}$Ni system. To explore this issue, we subdivide the experimental yields by their neutron-excess, and plot even and odd values separately. 
\begin{equation}
\label{eq:ynz}
Y_{N-Z}(Z)=Y(A,Z)/\sum_ZY(A,Z)\;,
\end{equation}
\noindent
which gives the yields of an isotope $Z$ with neutron excess $N-Z$.
Based on Eq.\ (\ref{eq:ysmoothFluct}), the staggering effect should be absent for odd values of $N-Z$ and $Y_{N-Z}(Z)$ will be smooth for $N-Z$ odd, since $Y(A,Z)$ becomes $Y_{\rm smooth}(A,Z)$, whereas important staggering effects should be expected when $N-Z$ is even due to the multiplying factor $e^{(-1)^Z\frac{\delta_p}{A^\tau}\frac{1}{T}}$.

The ratio:
\begin{equation}
\label{eq:ratio}
R_{N-Z}(Z)=Y_{N-Z}(Z) / \overline{Y}_{N-Z}(Z)
\end{equation}

\noindent
is plotted in the bottom panels of Fig.\ \ref{fig:58NiBE} for different values of neutron excess $N-Z=1,3,5$ (left panels) and $N-Z=0,2,4$ (right panels), obtained from the primary yields of the ISMM.
For succinctness, we focus on the $^{58}$Ni projectile, but similar results are also obtained with the other projectiles.
As expected from  Eq.\ (\ref{eq:ysmoothFluct}), the primary yields from ISMM clearly show that the OES is not obvserved in the case of odd neutron excess values whereas the OES is observed for even $N-Z$ values and the staggering diminishes as $Z$ increases. Aside from the different behavior for odd and even values of $N-Z$, the staggering effects seem to be independent of the $N-Z$ values. This behavior is consistent with the empirical binding energies, as illustrated in the top panels of Fig. \ref{fig:58NiBE}, where we plot the same type of ratio as above, but using the binding energies:
\begin{equation}
\label{eq:BEratio}
R(BE)_{N-Z}(Z)=BE_{N-Z}(Z) /~ \overline{BE}_{N-Z}(Z)
\end{equation}
Fig. \ref{fig:58NiBE} shows that staggering in the yields of fragments with odd neutron excess is not related to the primary phase of multifragmentation. 

\begin{figure}[tbh]
\includegraphics[width=8.5cm,angle=0]{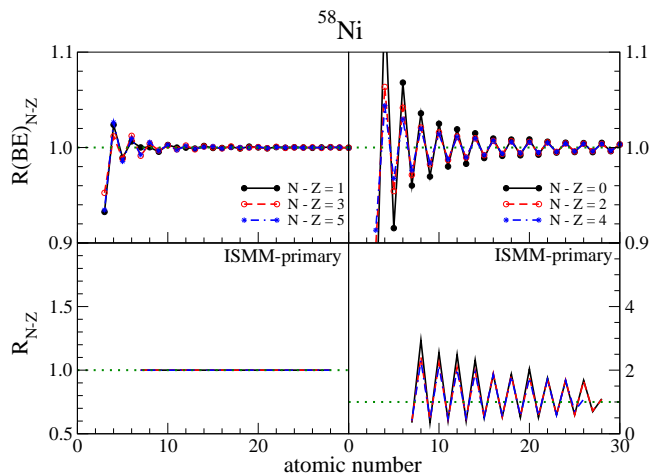}
\caption{\label{fig:58NiBE} (Color online) Comparison between staggering behavior in experimental binding energies (top panels) and primary yields from the ISMM (bottom panels). Fragments with odd neutron excess are shown on the left panels, while fragments with even neutron excess are shown on the right. }
\end{figure}

\begin{figure}[tbh]
\includegraphics[width=8.5cm,angle=0]{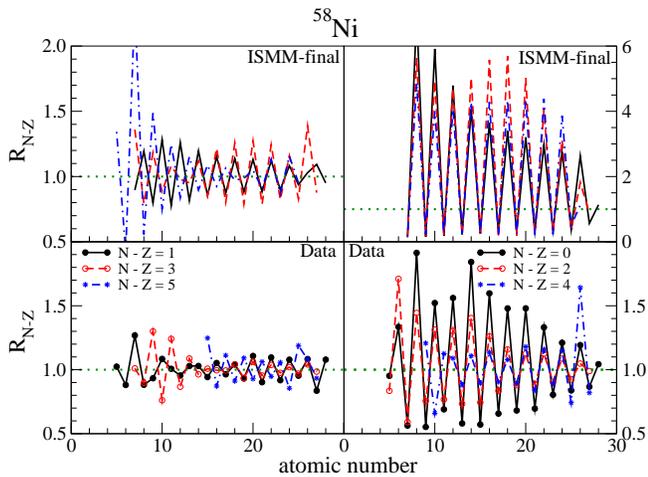}
\caption{\label{fig:58Ni} (Color online) The results obtained using the ISMM with Weisskopf-Ewing de-excitation are displayed in the top panels,
and the experimental results obtained with the projectile fragmentation data reported in Ref.\ \cite{projFragMSU2006} are shown in the bottom panels. 
Odd values of neutron excess are exhibited in the left panels, whereas even values are displayed in the right.
For details, see the text. }
\end{figure}
Since most of the hot primary  fragments have already decayed prior to being observed at the detectors, we use the implementation of the Weisskopf-Ewing statistical emission presented in Refs.\ \cite{ismmFlow2007,isoSMMTF} to estimate the effects of the de-excitation process on the fragment yields. The corresponding ratios are exhibited in the top panels of Fig.\ \ref{fig:58Ni}.
One notes that the fragments' de-excitation enhances the deviations from the average values so that $R_{N-Z}$ now exhibits fluctuations on the same order of magnitude as those observed in the experimental data for $N-Z$ odd.
However, the fluctuations are much larger in the case of even values of neutron excess, about twice the experimental values.
One should note that the scale used in the panels corresponding to the model results on the right hand side of Fig.\ \ref{fig:58Ni} is different from that adopted in the other panels of the figures.
Thus, although in the framework of the ISMM the de-excitation of the primary fragments is absolutely necessary to reproduce the experimentally observed staggering effects in the case of $N-Z$ odd, it leads to fluctuations which are much larger than those actually observed in the data for fragments with $N-Z$ even.
Therefore, it strongly suggests that $R_{N-Z}$ is very sensitive to the de-excitation scheme used in the model calculations and this observable could be used to constrain the treatment for the decay of excited fragments.

To examine if the reduced staggering for neutron rich nuclei $(N-Z=5)$ is also observed in the isotopes with odd values of neutron excess, $N-Z$, we turn to the data obtained from the neutron rich projectile, $^{64}$Ni. As shown in the top panels of Fig. \ref{fig:enhancement}, $R_5$ (joined by dashed lines) exhibits an exact opposite trend of the normal even-odd staggering effects; the cross-sections of odd-Z nuclei are enhanced relative to those with even-Z. The flip in the odd-even staggering for $N-Z=5$ isotopes can also be seen in the $^{58}$Ni projectile data plotted in the bottom left panel of Fig. \ref{fig:58Ni}. In addition, in contrast to the trend exhibited in even $N-Z$ isotopes where the staggering effects decrease with neutron richness, the magnitude of $R_5$ is much larger than $R_1$, for the lighter elements $(Z<14)$. The magnitude of the staggering effects in $R_5$ decreases and becomes similar to $R_1$ at higher $Z$. 

This behavior cannot be easily explained by the ISMM since the primary yields lead to $R_{N-Z}=1$ for $N-Z$ odd, as shown in the bottom left panel of Fig. \ref{fig:58NiBE}. In the framework of this model, it can only be explained by intricate correlations associated with the de-excitation of the primary fragments, as one sees in the top right panel of Fig. \ref{fig:enhancement}, which shows $R_{N-Z}$ computed with the final yields. However, the reversed OES takes place only at small $Z$ values, while it is observed over the whole $Z$ region.

One striking feature observed in the experimental data is the amplification of the staggering in the distribution associated with neutron deficient fragments ($N-Z=-1$) compared with the ratios obtained with neutron rich ones ($N-Z=1$).
The corresponding experimental observations are shown in the bottom left panel of Fig.\ \ref{fig:enhancement}, for the $^{58}$Ni projectile.
This amplification has no explanation in the framework of the ISMM since Eq.\ (\ref{eq:ysmoothFluct}) predicts that the primary yields would be strictly smooth for $N-Z=\pm 1$.
To check whether the de-excitation treatment employed in this work could account for the observed enhancement, the ratios obtained with the final ISMM yields are displayed in the bottom right panel of Fig.\ \ref{fig:enhancement}.
This enhancement is not seen in the model calculations. Clearly, more detailed investigations are needed to understand this feature.

The theoretical model employed in this work could not quantitatively explain many features observed experimentally. Further theoretical investigation might lead to an improvement in the understanding of the de-excitation process. 
For example, this model currently assumes that the pairing term in the binding energy has no temperature dependence, a dependence which would affect how the hot fragments de-excite.

\begin{figure}[tbh]
\includegraphics[width=8.5cm,angle=0]{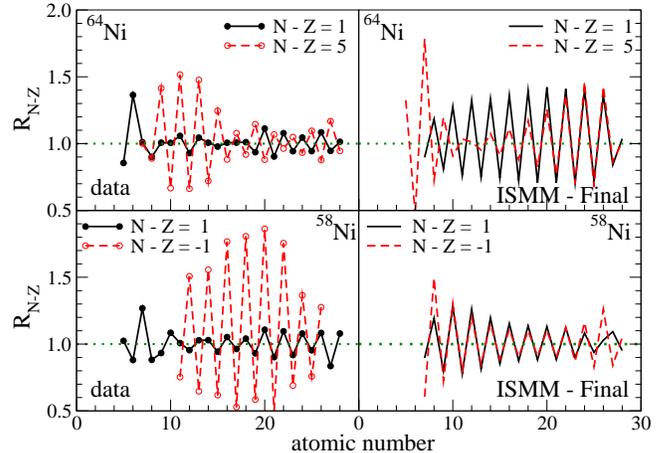}
\caption{\label{fig:enhancement} (Color online) Comparison of the ratios obtained with $N-Z=-1$  and $N-Z=1$ for the $^{58}$Ni projectile (lower panels) and with $N-Z=1$ and $N-Z=5$ in the case of the $^{64}$Ni projectile (upper panels).
The experimental ratios are calculated using the yields of Ref.\ \cite{projFragMSU2006}.
For details, see the text.}
\end{figure}

\end{section}

\begin{section}{Concluding remarks}
\label{sect:conclusions}
We have examined the effects of the de-excitation on the staggering observed in the yields of the fragments produced in projectile fragmentation.
The model qualitatively explains this feature through odd-even effects associated with the pairing term of the nuclear binding energy.
However, we have found that the ISMM is not able to reproduce the experimentally observed odd-even staggering effects in the case of fragments with odd $N-Z$, if the effects associated with the de-excitation of the hot primary fragments are not taken into account.
Even though the sequential decay fixes this shortcoming of the model, it leads to fluctuations which are much larger than observations in the case of fragments with $N-Z$ even.
Further examination of the staggering effect exposes additional deficiencies of the models employed in this work, as the enhancement of the ratio obtained with neutron deficient ($N-Z=-1$) fragments compared to that calculated with neutron rich ($N-Z=1$) nuclei cannot be reproduced theoretically.
Moreover, the phase shift in the staggering behavior for $N-Z=1$ and $N-Z=5$ leads to a flip at large $Z$. This is only partially reproduced by the model, which predicts a flip at small $Z$.
These findings reveal that the observed staggering is quite sensitive to the treatment employed in describing the decay of the primordial hot fragments and can therefore be used to constrain these treatments.

\end{section}

\begin{acknowledgments}
We would like to acknowledge CNPq,  FAPERJ BBP grant, FAPERGS and the joint PRONEX initiatives of CNPq/FAPERJ under
Contract No.\ 26-111.443/2010, for partial financial support.
This work is supported by the US National Science Foundation under Grant No. PHY-1102511 and No. PHY-0822648.
\end{acknowledgments}

\bibliography{manuscript}
\bibliographystyle{apsrev4-1}

\end{document}